\documentclass[aip,showpacs,superscriptaddress,twocolumn,reprint]{revtex4}%

\usepackage{amsfonts}
\usepackage{amsmath}
\usepackage{amssymb}
\usepackage{graphicx}%
\begin{document}
\title{Thermophysical properties for shock compressed polystyrene}
\author{Cong Wang}
\affiliation{LCP, Institute of Applied Physics and Computational Mathematics, P.O. Box
8009, Beijing 100088, People's Republic of China}
\author{Xian-Tu He}
\affiliation{LCP, Institute of Applied Physics and Computational Mathematics, P.O. Box
8009, Beijing 100088, People's Republic of China}
\affiliation{Center for Applied Physics and Technology, Peking University, Beijing 100871,
People's Republic of China}
\author{Ping Zhang}
\thanks{Corresponding author: zhang\underline{ }ping@iapcm.ac.cn}
\affiliation{LCP, Institute of Applied Physics and Computational Mathematics, P.O. Box
8009, Beijing 100088, People's Republic of China}
\affiliation{Center for Applied Physics and Technology, Peking University, Beijing 100871,
People's Republic of China}
\affiliation{National Institute for Fusion Science, Toki-shi 509-5292, Japan}

\begin{abstract}
We have performed quantum molecular dynamic simulations for warm dense
polystyrene at high pressures. The principal Hugoniot up to 790 GPa is derived
from wide range equation of states, where contributions from atomic
ionizations are semiclassically determined. The optical conductivity is
calculated via the Kubo-Greenwood formula, from which the dc electrical
conductivity and optical reflectivity are determined. The nonmetal-to-metal
transition is identified by gradual decomposition of the polymer. Our results
show good agreement with recent high precision laser-driven experiments.

\end{abstract}

\maketitle

In high energy density ($E/V\mathtt{\geq}10^{11}$J/m$^{3}$) physics (HEDP),
pressure-induced response of materials, which can be probed through shock wave
experiments
\cite{PBX:Knudson:2009,PBX:Nellis:2006,PBX:Hicks:2008,PBX:Hicks:2009}, is of
crucial interest for inertial confinement fusion (ICF). Fundamental
experiments in typical ICF designs, such as the investigation of hydrodynamic
instabilities, have introduced plastic compounds as basic candidates to
achieve high gain \cite{PBX:Lindl:1998}. Furthermore, in recent fast ignitor
experiments, the electrical properties of polymers have been often implicated
for the transport of relativistic electron beams in solid targets as compared
to conducting materials \cite{PBX:Cai:2009}. Understanding the behavior of
polymers at several megabar regime brings better insight into physical
processes in ICF.

Some of the ablators in the indirect-drive mode of National Ignition Facility
(NIF), will be made of glow-discharge polymer (GDP) with various levels of
germanium doping (Ge-GDP) \cite{PBX:Haan:2007}. Due to the fact that no high
pressure data exist for Ge-GDP, polystyrene (CH), which is closest in
structure, is considered as a coarse indicator for shock timing simulations of
NIF targets involving such ablators \cite{PBX:Barrios:2010}. The absolute
equation of states (EOS) of polystyrene along the principal Hugoniot curve has
been studied by using gas gun up to 50 GPa
\cite{PBX:LASL:1980,PBX:McQueen:1970,PBX:Thiel:1977,PBX:Bushman:1996} and high
energy laser-driven shock waves up to 4000 GPa
\cite{PBX:Cauble:1997,PBX:Ozaki:2005,PBX:Ozaki:2009}. High energy dynamic
compressions produce the so-called warm dense matter (WDM), where simultaneous
dissociations, ionizations, and degenerations make the transition between
condensed matter physics to plasma physics. The EOS and the relative
properties, such as the Hugoniot curve, are important features in this
context. Moreover, electrical properties such as the dielectric function are
closely related to the dynamic conductivity, which determines the dc
electrical conductivity in the static limit. Then optical reflectance can also
be extracted. All these quantities are indispensable in characterizing the
unique behavior of shock compressed polystyrene, especially for high pressures
at, or exceeding, 1 Mbar. Recently, the development and application of quantum
molecular dynamic (QMD) simulation \cite{PBX:Kietzmann:2008} techniques for
WDM are available due to the enormous progress in computer capacity. QMD
simulations, where quantum effects and correlations are systematically
treated, have conducted highly predictive results to describe WDM.

In the present work, QMD simulations are applied to calculate a broad spectrum
of thermophysical properties for shock compressed polystyrene. The
self-consistent electronic structure calculation within density functional
theory (DFT) yields the charge density distribution in the simulation
supercell at every time step. The ion-ion pair correlation function (PCF),
which is important for characterizing and identifying phase transitions, can
be given by molecular dynamics run. The Hugoniot curve, which is derived from
EOS data for a wide region of densities and temperatures, is determined and
compared with available experimental and theoretical results. As a starting
point, we use the Kubo-Greenwood formula to evaluate the dynamic conductivity
$\sigma(\omega)$, from which the dc conductivity, the dielectric function
$\epsilon(\omega)$, and the optical reflectivity can be settled.

We introduce \textit{ab} initio plane wave code VASP
\cite{PBX:Kresse:1993,PBX:Kresse:1996} to perform QMD simulations. The
elements of our calculations consist of a series of volume-fixed supercells
including $N$ atoms, which are repeated periodically throughout the space. By
involving Born-Oppenheimer approximation, electrons are quantum mechanically
treated through plane-wave, finite-temperature DFT \cite{PBX:Recoules:2009},
where the electronic states are populated according to Fermi-Dirac
distributions. Sufficient occupational bands are included in the overall
calculations (the occupational number down to 10$^{-6}$ for electronic states
are considered). The exchange-correlation functional is determined by
generalized gradient approximation (GGA) with the parametrization of
Perdew-Wang 91 \cite{PBX:Perdew:1991}. The ion-electron interactions are
represented by a projector augmented wave (PAW) pseudopotential
\cite{PBX:Blochl:1994}. Isokinetic ensemble (NVT) is adopted in the present
simulations, where the ionic temperature $T_{i}$ is controlled by No\'{s}e
thermostat \cite{{PBX:Nose:1984}}, and the system is kept in local equilibrium
by setting the electron ($T_{e}$) and ion ($T_{i}$) temperatures to be equal.

\begin{figure}[ptb]
\centering
\includegraphics[width=8.0cm]{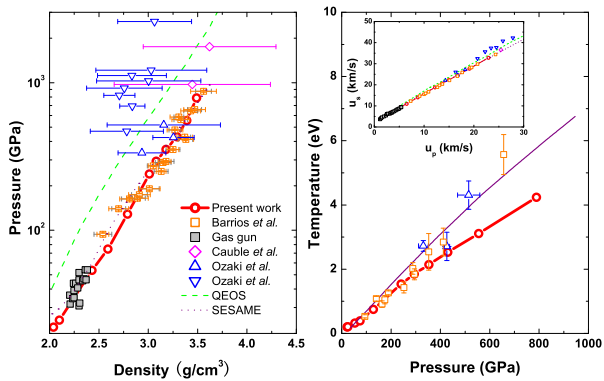} \caption{(Color online) The $P$-$V$
(left panel) and $T$-$P$ (right panel) Hugoniot curves of shocked polystyrene.
Inset is the ($u_{s}$, $u_{p}$) diagram. Previous results are also shown for
comparison. Experiments: gas-gun results (the gray filled squares);
\cite{PBX:LASL:1980,PBX:McQueen:1970,PBX:Thiel:1977,PBX:Bushman:1996} absolute
measurements on NOVA (magenta diamonds); \cite{PBX:Cauble:1997} high energy
laser-driven measurements by Ozaki \emph{et al.}
\cite{PBX:Ozaki:2005,PBX:Ozaki:2009} (blue upward and downward triangles) and
Barrios \emph{et al.} (orange squares). \cite{PBX:Barrios:2010} Theories: QEOS
and SESAME \cite{PBX:Lyon:1998} results are denoted by green dashed line and
purple dotted line, respectively.}%
\label{fig_hugoniot}%
\end{figure}

The simulations have been done for 128 atoms (namely, eight C$_{8}$H$_{8}$
units) in a supercell. A plane-wave cutoff energy of 600.0 eV is selected, so
that a convergence of better than 3\% is secured. The Brillouin zone is
sampled by $\Gamma$ point and 3$\times$3$\times$8 Monkhorst-Pack scheme
\cite{PBX:Monkhorst:1976} \textbf{k} points in molecular dynamic simulations
and electronic structure calculations, respectively, because EOS
(conductivity) can only be modified within 5\% (15\%) for the selection of
higher number of \textbf{k} points. The densities adopted in our simulations
range from 1.05 g/cm$^{3}$ to 3.50 g/cm$^{3}$ and temperatures between 300 and
50000 K, which highlight the regime of principal Hugoniot. All the dynamic
simulations are lasted 6000 steps, and the time steps for the integrations of
atomic motion are selected according to different densities (temperatures)
\cite{PBX:timestep}. Then, the subsequent 1000 step simulations are used to
calculate EOS as running averages.

\begin{table}[!htbp]
\centering \caption{Polystyrene Hugoniot results from QMD
simulations. }
\begin{tabular}{cccccccccccccccccccccccccccccccccccccccc}
\hline\hline $\rho$ (g/cm$^{3}$)& $P$ (GPa) & $T$ (eV) & $u_{s}$ (km/s) & $u_{p}$ (km/s)\\
\hline
2.04 & 22.06   & 0.20 & 6.59      & 3.19    \\
2.10 & 24.61   & 0.21 & 6.85      & 3.42    \\
2.43 & 53.48   & 0.32 & 9.48      & 5.37    \\
2.59 & 74.54   & 0.39 & 10.93     & 6.49    \\
2.79 & 128.56  & 0.75 & 14.01     & 8.74    \\
3.01 & 240.46  & 1.54 & 18.76     & 12.21   \\
3.08 & 293.98  & 1.85 & 20.60     & 13.59   \\
3.18 & 352.98  & 2.14 & 22.40     & 15.01   \\
3.29 & 429.62  & 2.53 & 24.51     & 16.69   \\
3.39 & 554.96  & 3.11 & 27.67     & 19.10   \\
3.49 & 789.07  & 4.24 & 32.79     & 22.92   \\
\hline\hline
\end{tabular}
\label{H_data}
\end{table}

The nature of high pressure behavior for warm dense polystyrene is dominated
by a two-stage transition, that is, dissociations and ionizations. QMD
simulations have been demonstrated to be powerful in describing as well as
understanding chemical decompositions and electronic excitations of shock
compressed materials, namely, the shocked atomic and electronic structures can
be effectively treated within QMD. Importantly, however, we should address
that the calculated EOS data from direct QMD simulations on shocked
polystyrene should be corrected due to the following aspects: (i) At
laser-driven temperatures as high as of 5$\mathtt{\sim}$6 eV ($\mathtt{\sim}%
$10 Mbar), the considerable effect of atomic ionizations on EOS should be
appropriately included beyond QMD; (ii) QMD energy consists of kinetic and
potential energy (for both ions and electrons), while, the well known
zero-point vibration energy (ZPVE), van der Waals energy (VDWE), and so as the
demanded energy for phase transitions, which are particularly significant in
WDM, are excluded; (iii) The pressure from QMD only describes the interaction
contributions, and the ideal part for noninteracting particles are missed. To
overcome these obstacles, in our present calculations for polystyrene, ZPVE,
which is important for determining the low-pressure internal energy for the
reference state along the Hugoniot curve, is simply added onto the QMD output.
Condensed matter to plasma transition energy mainly comes from atomic
ionization energy, which is also added onto QMD data, and the reference
ionization degree is semiclassically determined \cite{PBX:Wang:2010b}. VDWE
has been treated as small as negligible. As for another important
parameter---pressure, contributions from both interacting and noninteracting
parts (ions and free electrons, separately) are included in the present EOS data.

High precision EOS data are essential for understanding the electrical and
optical properties of polystyrene under extreme conditions. The present EOS
have been examined theoretically through the Rankine-Hugoniot (RH) equations,
which follow from conservation of mass, momentum, and energy across the front
of shock waves. The locus of points in ($E,P,V$)-space described by RH
equations satisfy the following relations:
\begin{equation}
E_{1}-E_{0}=\frac{1}{2}(P_{1}+P_{0})(V-V_{0}),\label{equ_hugoniot1}%
\end{equation}%
\begin{equation}
P_{1}-P_{0}=\rho_{0}u_{s}u_{p},\label{equ_hugoniot2}%
\end{equation}%
\begin{equation}
V_{1}=V_{0}(1-u_{p}/u_{s}),\label{equ_hugoniot3}%
\end{equation}
where subscripts 0 and 1 present the initial and shocked state, and $E$, $P$,
$V$ denote internal energy, pressure, volume, respectively. The $u_{s}$ and
$u_{p}$ correspond to the shock and mass velocities of the material behind the
shock front. As the starting point along the Hugoniot, the initial density is
$\rho_{0}$=1.05 g/cm$^{3}$, and the internal energy is $E_{0}$=$-$87.44 kJ/g
at a temperature of 300 K. Compared to the high pressure of shocked states
along the Hugoniot, the initial pressure $P_{0}$ can be treated approximately
as zero. We use smooth functions to fit the internal energy and pressure in
terms of temperature at sampled density, and derive Hugoniot point from Eq.
\ref{equ_hugoniot1}. The calculated Hugoniot data are listed in Table
\ref{H_data}.

The principal Hugoniot curve is plotted in Fig. \ref{fig_hugoniot}, where
previous theoretical and experimental results are also shown for comparison.
The EOS of polystyrene has been previously probed by gas gun experiments,
which have the advantage of high precision, but the pressure hardly exceeds 50
GPa. Recently, high energy laser-driven experiments have detected pressures up
to 4000 GPa. Meanwhile, however, large error bars were introduced at above 100
GPa \cite{PBX:Cauble:1997,PBX:Ozaki:2005}, and thus the use of low-precision
EOS of the polymer to predict the behavior of NIF Ge-GDP ablator materials
provides an unacceptable uncertainty. Then, high precision EOS (up to 1000
GPa) for polystyrene was obtained by using $\alpha$-quartz as an
impedance-matching (IM) standard \cite{PBX:Barrios:2010}. As shown in Fig.
\ref{fig_hugoniot}, previous data by Ozaki \emph{et al. }\cite{PBX:Ozaki:2005}
(IM with an aluminum standard) and Cauble \emph{et al. }\cite{PBX:Cauble:1997}
(absolute data) show clearly stiffer behavior compared to those results with
quartz standard \cite{PBX:Ozaki:2009,PBX:Barrios:2010} and SESAME model
\cite{PBX:Lyon:1998}. The possible reason for this stiff behavior is likely
due to x-ray preheating of the samples, as has been stated by those authors.
Then, thicker pushers and low-Z ablators were used in the newly measured data
from Ozaki \emph{et al. }\cite{PBX:Ozaki:2009} to reduce pre-heating of the
samples. The experiments, where IM with a quartz standard were also used, show
results that are closer to those by Barrios \emph{et al.}. In our QMD
simulations, we find that direct calculated EOS without corrections mentioned
above are only valid in the low pressure ($P\mathtt{<}$100 GPa and
$T\mathtt{<}$1 eV) regime, beyond which no Hugoniot points can be found from
the pure QMD data. For higher pressures, due to our introduction of
corrections to direct QMD results, the calculated principal Hugoniot curve are
greatly softened, and the results show good agreements with those experiments
where quartz standard was used.

\begin{figure}[ptb]
\centering
\includegraphics[width=8.0cm]{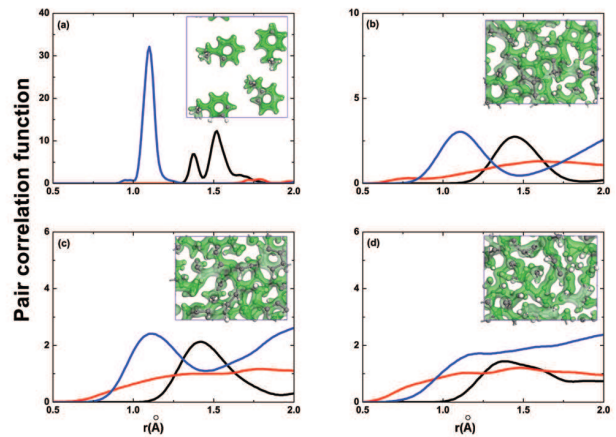}\caption{(Color online) Pair
correlation functions for four different points along the principal Hugoniot
curve. (a) $\sim$ (d) correspond to the starting point, 53.48 GPa, 74.54GPa,
and 128.56 GPa, respectively. C-H, C-C, and H-H bonds are denoted by blue,
black, and red lines, respectively. Insets are the sampled atomic structure
and charge density distribution at each state (gray balls for carbon and white
balls for hydrogen). }%
\label{fig_rdf}%
\end{figure}

Temperature, as has been focused as one of the most important parameters in
experiments, is difficult to be measured because of the uncertainty in
determining the optical-intensity loss for ultraviolet part of the spectrum in
adiabatic or isentropic shock compressions, especially for the temperature
exceeding several eV \cite{Chen2010}. QMD simulations can provide efficient
predictions for shock temperatures. As has been shown in Fig.
\ref{fig_hugoniot} (right panel), our calculated Hugoniot temperatures are
accordant with experimental results up to 2 eV, and discrepancy emerges at
higher temperatures. Beyond 2 eV, the predicted temperatures from both
experiments and SESAME model at a given pressure are higher compared to our calculations.

We also examine the structural transition of polystyrene under shock
compressions by using PCF, which is evaluated at equilibrium during the
molecular dynamics simulations. Along the Hugoniot, four points of PCF are
shown in Fig. \ref{fig_rdf}. At the initial state [see Fig. \ref{fig_rdf}(a)],
the ideal condensed polymer phase is indicated by one peak for C-H bond,
meanwhile, two peaks ($\pi$-type and $sp^{3}$-type) for C-C bond. With the
increase of pressure ($\sim$20 GPa), phenyl decomposes, which is indicated by
that the two C-C peaks merge together (a transverse from $\pi$ bond to
$sp^{2}$, $sp^{3}$ like bond). At this stage diamondlike carbon nanoparticles
(with defects) are formed. The present phenomenon agrees well with the
behavior of shock compressed benzene \cite{PBX:wang:2010c,PBX:Nellis:2001}.
Further increase of pressure produces molecular and atomic hydrogen, which is
indicated by the peak in H-H PCF around 0.75 \r{A}\ [Fig. \ref{fig_rdf}(b)].
Decomposition of hydrocarbons continues until 128 GPa, where PCF shows no
obvious evidence for the existence of C-H bond. Whereas, for higher pressures,
the PCFs do not show visible difference, and are thus not presented here.
SESAME model suggested that the softening behavior of Hugoniot is caused by
the chemical decomposition of the polymer at 200 $\sim$ 400 GPa. Whereas, our
calculations show that the decomposition pressure lies between 20 and 128 GPa,
which is much lower than that of SESAME. Note that the chemical decompositions
do not contribute much to the softening behavior, but the ionizations really do.

\begin{figure}[ptb]
\centering
\includegraphics[width=8.0cm]{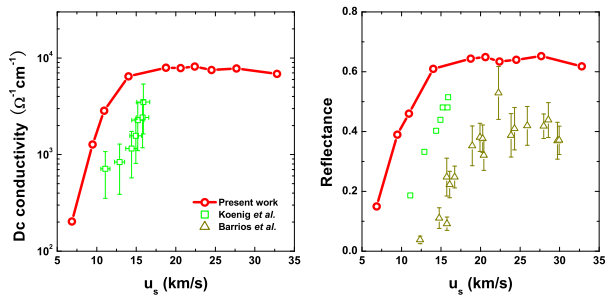}\caption{(Color online) Dc conductivity
(left panel) and optical reflectivity at the wavelength of 532 nm (right
panel). Previous measurements by Koenig \emph{et al.} (green squares)
\cite{PBX:Koenig:2003} and Barrios \emph{et al.} (upward triangles)
\cite{PBX:Barrios:2010} are also plotted.}%
\label{fig_dc}%
\end{figure}

Having clarified the EOS, let us turn now to study the electrical and optical
properties for shocked polystyrene. Based on Kubo-Greenwood formula, the
electrical and optical properties can be obtained as described in Ref.
\cite{PBX:wang:2010c}. In order to get converged results, 30 independent
snapshots, which are selected during one molecular dynamics simulation at
given conditions, are picked up to calculate dynamic conductivity as running
averages. The dc conductivity $\sigma_{dc}$, which follows from the static
limit $\omega\mathtt{\rightarrow}0$ of $\sigma_{1}(\omega)$ is then evaluated
and plotted in Fig. \ref{fig_dc} (left panel) as a function of shock velocity.
Initially, $\sigma_{dc}$ increases rapidly with shock velocity up to 14 km/s
(128 GPa) towards the formation of metallic state of polystyrene. Then, one
can find from Fig. \ref{fig_dc} that $\sigma_{dc}$ keeps almost invariant and
the shock compressed polystyrene maintains its metallic behavior. The onset of
metallization ($\sigma_{dc}\mathtt{>}10^{3}\Omega^{-1}$cm$^{-1}$) is observed
at around 10 km/s ($\mathtt{\sim}$50 GPa), where dissociation of the polymer
and formations of hydrogen (molecular and atomic species) govern the
characteristics of warm dense polystyrene. We stress here that the nonmetal to
metal transition is induced by gradual chemical decompositions and thermal
activations of the electronic state, instead of atomic ionization, which is
not observed until 128 GPa with respect to the charge density distribution in
the QMD simulations. In the same pressure range (20$\mathtt{\sim}$200 GPa), we
observe larger $\sigma_{dc}$ compared to experimental measurement
\cite{PBX:Koenig:2003}. This overestimation of $\sigma_{dc}$ values could be
attributed to the use of DFT-based molecular dynamics, which is known to
underestimate band gaps in many systems.

Optical reflectance, from which emissivity can be derived, is of great
interest in experiment for determining shock temperature. Along the Hugoniot,
we show in Fig. \ref{fig_dc} (right panel) the variation of optical
reflectance at 532 nm as a function of shock velocity. Optical properties of
polystyrene have been experimentally studied by Koenig \emph{et al.
}\cite{PBX:Koenig:2003} with $u_{s}$ of 11$\mathtt{\sim}$16 km/s
(80$\mathtt{\sim}$170 GPa), and steadily increasing reflectivities reaching
values up to $\mathtt{\sim}$50 \% were observed. The results by Ozaki \emph{et
al.} indicated reflectivity from 16\% to 42\% in $u_{s}$ range of
22$\mathtt{\sim}$27 km/s (300$\mathtt{\sim}$500 GPa) \cite{PBX:Ozaki:2009}.
Recent experiments by Barrios \emph{et al. }\cite{PBX:Barrios:2010} suggested
a drastic increase in reflectivity around the Hugoniot pressure of 100 GPa,
followed by saturated value of 40\% around 250$\mathtt{\sim}$300 GPa.
Discrepancies in reflectivities among experiments have been claimed to be
arisen from probe-beam stability or from differences in diagnostic
configurations. Along the Hugoniot, our QMD calculations provide the general
feature for the optical reflectance---steep increase (from 15\% $\mathtt{to}$
60 \%) followed by saturation (at $\sim$128 GPa). The change in optical
reflectivity with pressure can be interpreted as the gradual
insulator-conductor transition at above 20 GPa, where the atoms strongly
fluctuate with neighbors to dense, partially ionized plasma at high pressures
above 128 GPa.

In summary, we have carried out QMD simulations to study the thermophysical
properties for warm dense polystyrene. The Hugoniot EOS data of polystyrene up
to $\mathtt{\sim}$790 GPa, which is in agreement with dynamic experiments in a
wide range of shock conditions, has been evaluated through QMD calculations
and corrected by taking into account the atomic ionization. A two-stage
(dissociation and ionization) transition governs the characteristic of
polystyrene under extreme conditions. Contribution from chemical decomposition
demonstrates the steep increase of $\sigma_{dc}$ and optical reflectance
observed at 20$\mathtt{\sim}$128 GPa. While, soften behavior of the Hugoniot
is dominated by atomic ionizations for higher pressure.

This work was supported by NSFC under Grants No. 11005012 and No. 51071032, by
the National Basic Security Research Program of China, by the National
High-Tech ICF Committee of China, and by the Core University Program.


\begin{thebibliography}{99}
\bibitem {PBX:timestep}The time steps have been taken as $\Delta t$%
=$a/20\sqrt{k_{B}T/m}$, where $a$=$(3/4\pi n_{i})^{1/3}$ is the ionic sphere
radius ($n_{i}$ is the ionic number density), $k_{B}T$ presents the kinetic
energy, and $m$ is the ionic mass.
\end{thebibliography}

\end{document}